\documentstyle[aps,prb,eqsecnum,epsf]{revtex}
\newcommand{\be}{\begin{equation}}
\newcommand{\ee}{\end{equation}}
\newcommand{\bea}{\begin{eqnarray}}
\newcommand{\eea}{\end{eqnarray}}

\def\b2c{\begin{multicols}{2}}
\def\e2c{\end{multicols}}

\def\doublespacing{%
    \def\default@spacing{\baselineskip=23.5pt plus .5pt minus .2pt}}
\def\onehalfspacing{%
    \def\default@spacing{\baselineskip=19.5pt plus .5pt minus .2pt}}
\def\singlespacing{%
    \def\default@spacing{\baselineskip=15.5pt plus .5pt minus .2pt}}

\begin{document}
\title{Quantum Mechanics of Extended Objects}
\author{Ramchander R. Sastry}
\address{Center for Particle Physics \\
University of Texas at Austin \\
  Austin, Texas 78712-1081.}
\date{\today}
\maketitle
\large %
\baselineskip=22.5pt plus .5pt minus .2pt
\begin{abstract}
 We propose a quantum mechanics of extended objects
that accounts for the finite extent of a particle defined via its
Compton wavelength. The Hilbert space representation theory of
such a quantum mechanics is presented and this representation is
used to demonstrate the quantization of spacetime. The quantum
mechanics of extended objects is then applied to two paradigm
examples, namely, the fuzzy (extended object) harmonic oscillator
and the Yukawa potential. In the second example, we theoretically
predict the phenomenological coupling constant of the $\omega$
meson, which mediates the short range and repulsive nucleon force,
as well as the repulsive core radius.
\\
\end{abstract}
\normalsize %
\baselineskip=19.5pt plus .5pt minus .2pt
\section{Introduction}
 The representation of a particle as an
idealized point has long been used in physics. In fact, this
representation is central to classical mechanics and serves us
well even in quantum mechanics. In this paper we adopt a viewpoint
in which the finite extent or fuzziness of a particle is taken
into consideration thereby treating the particle as an extended
object.  Such a treatment becomes important and necessary when the
confines of the quantum system in which the particle is placed
becomes comparable to the finite extent of the particle.  The
finite extent or fuzziness of a particle is quantified via its
Compton wavelength which can be defined as the lower limit on how
well a particle can be localized.  In nonrelativistic quantum
mechanics, the lower limit is zero since we admit position
eigenkets $|x\rangle$.  But in reality, as we try to locate the
particle with greater accuracy we use more energetic probes, say
photons to be specific. To locate a particle to some $\Delta x$ we
need a photon of momentum
\be
\Delta p \approx \frac{\hbar}{\Delta x}.
\ee
The corresponding energy of the photon is
\be
\Delta E \approx \frac{\hbar c}{\Delta x}.
\ee
If this energy exceeds twice the rest energy of the particle,
relativity allows the production of a particle-antiparticle
pair in the measurement process.  So we demand
\be
\frac{\hbar c}{\Delta x} \leq 2mc^{2}
\quad \mbox{or} \quad
\Delta x \geq \frac{\hbar}{2mc} \approx \frac{\hbar}{mc}.
\ee
Any attempt to further localize the particle will lead to
pair creation and we will have three (or more) particles
instead of the one we started to locate.  Therefore, the
Compton wavelength of a particle measures the distance over
which quantum effects can persist  The point particle
approximation used in nonrelativistic quantum mechanics
suffices to describe the dynamics since the confines of
the quantum systems under consideration are much larger
than the finite extent of the confined particles.  For example, in
the analysis of the hydrogen atom, the fuzziness or the
size of the electron is $\alpha$ times smaller than the
size of the atom $a_{0}$
\be
\frac{\hbar/mc}{a_{0}} = \alpha \approx \frac{1}{137}.
\ee
Thus, in the case of the hydrogen atom and in general, for
the quantum theory of atoms, the quantum mechanics of point
 particles gives an accurate description.

In this paper we develop the Hilbert space representation
theory of the quantum mechanics of extended objects.  We use
this representation to demonstrate the quantization of spacetime
following which we analyze two paradigm examples: fuzzy
harmonic oscillator and the Yukawa potential.  In the
second example, the quantum mechanics of extended objects
enables us to predict the phenomenological coupling
constant of the $\omega$ meson as well as the radius of
the repulsive nucleon core.

\section{Quantum Mechanics of Extended Objects}
  We have established the necessity for taking into consideration
the nonzero size of a particle.  In order to incorporate the
fuzziness or size of a particle into our dynamics we
introduce the following representation for position and
momentum in one dimension in units where $\hbar = c = 1$.
For position space,
\bea
X_{f} & = & (Xe^{-P^{2}/m^{2}})
\rightarrow (xe^{-P^{2}/m^{2}}) \nonumber \\
P & \rightarrow & {-i}\frac{d}{dx}  \\
\left[ X_f, P \right] & = & i e^{-P^{2}/m^{2}}, \nonumber
\eea
and for momentum space,
\bea
X_{f} & = & e^{-P^{2}/2m^{2}}Xe^{-P^{2}/2m^{2}}
\rightarrow i e^{-P^{2}/2m^{2}}\frac{d}{dp}e^{-P^{2}/2m^{2}}
\nonumber \\
P & \rightarrow & p \\
\left[X_{f},P \right] & = & i e^{-p^{2}/m^{2}}. \nonumber
\eea
where $(AB) \equiv {(AB + BA)}/{2}$. Symmetrization has also
been employed in the momentum space representation in order to
preserve the Hermiticity of the noncommuting fuzzy position
operator $X_{f}$.  In contradistinction to the quantum
mechanics of point particles where the position operator
has a smooth coordinate representation consisting of a sequence
of points, the fuzzy position operator is convolved with a
Gaussian in momentum space which has as its width the Compton
wavelength ${1}/{m}$.  The convolution with the Gaussian
has the effect of smearing out these points and in the limit
as the Compton wavelength vanishes we recover the standard
operator assignments of ordinary quantum mechanics.  For
simplicity, consider the effect of the fuzzy
position operator $X_{f}$ on an acceptable wavefunction in position space,
that is, one which is square integrable and has the right behavior
at infinity:
\bea
X_{f}\psi(x)  &=&  (xe^{-P^{2}/m^{2}}) \psi (x) \nonumber \\
 &=&  \frac{m}{4\sqrt{\pi}}\,\left[
\int_{-\infty}^{\infty} d\lambda\, x\,
e^{iP\lambda - m^{2}\lambda^{2}/4}\psi(x)\,+
\, \int_{-\infty}^{\infty} d\lambda\,
e^{iP\lambda - m^{2}\lambda^{2}/4}[x\psi(x)] \right]  \nonumber \\
& = & \frac{m}{4\sqrt{\pi}} \int_{-\infty}^{\infty}\,
d\lambda(x + \frac{\lambda}{2}) \psi(x + \lambda)
e^{-m^{2}\lambda^{2}/4}.
\eea
The translation of $\psi(x)$ by $\lambda$ and the subsequent
integration over all possible values of $\lambda$ weighted by
a Gaussian measure has the effect of smearing out the position.
The commutation relation obeyed by $X_{f}$ and $P$ is manifestly
noncanonical and does not depend on the representation.  A direct
consequence of this commutation relation is the uncertainty relation.
\be
\label{f-eqn}
\Delta X_{f}\Delta P\, \geq\, \frac{1}{2}|\langle e^{-P^{2}/m^{2}}
\rangle|.
\ee
Now, for any two observables $A$ and
$B$ which satisfy
$\left[A,B\right]|\psi\rangle = 0$ for some
nontrivial $|\psi\rangle$,
 with uncertainties $\Delta A$ and
$\Delta B$ such that
$|{\Delta A}/{\langle A\rangle}|\ll 1$ and
$|{\Delta B}/{\langle B\rangle}|\ll 1$, we have the relation
\be
\label{ab-eqn}
\Delta ((AB)) = \langle A\rangle \Delta B + \langle B\rangle \Delta A,
\ee
where again $(AB) \equiv {(AB + BA)}/{2}$.  The special case
$\left[A,B\right] = 0$ corresponds to compatible variables.
We observe that whenever simultaneous eigenkets exist
\bea
\langle AB\rangle &=& \int da\, db\, P(ab)\,ab = \int da\, db\,
P(a)P(b)\, ab  \nonumber \\
&=& \langle A\rangle \langle B\rangle
\eea
where $P(ab) = |\langle ab|\psi \rangle|^{2}$ and the proof
of Eq.~(\ref{ab-eqn}) follows.
 In our case,
\be
\left[X,e^{-P^{2}/m^{2}}\right]|\psi \rangle = 0
\mbox{ only if }|\psi\rangle = {\rm constant}.
\ee
Hence, there
exists at least one nontrivial simultaneous eigenket for which
$[X,e^{-P^{2}/m^{2}}]$ has a zero eigenvalue.
We can always choose this eigenket to establish the validity
of Eq.~(\ref{ab-eqn}) for our operators $X$ and
$e^{-P^{2}/m^{2}}$ along the lines shown above.
As a consequence, we obtain the modified uncertainty
principle (reinserting $\hbar$ for clarity)
\be
\Delta X\Delta P \,\geq\, \frac{\hbar}{2}
\,+\, \frac{2\langle X\rangle \langle P\rangle }{m^{2}}(\Delta P)^{2}.
\ee
The uncertainty product goes up because of the fuzziness
we have introduced in the position.  Consequently, there exists
a minimal uncertainty in position given by
\be
\Delta X_{0} = \frac{2}{m}\sqrt{\langle X\rangle\langle P\rangle\hbar}.
\ee
The existence of minimal uncertainties and their consequences for
structure were first examined by Kempf, albeit, in a different
context \cite{kempf1}\cite{kempf2}.  We note that the product
$\langle X\rangle \langle P\rangle$ is in general nonnegative. It
can be made negative by moving the center of coordinates but this
would imply that the Hamiltonian of the underlying system is
translationally invariant such as the free particle or the
particle in a box (for bound systems $\langle P\rangle = 0$). For
all such systems the Hamiltonian does not depend on the position
(or fuzzy position) and incorporating the fuzziness of the
particle into our quantum description is irrelevant to the
dynamics.   Hence, the Compton wavelength can be set to zero in
such cases which is the correspondence limit with ordinary quantum
mechanics.  If we view the uncertainty product as a measure of the
cell volume of phase space we observe that quantized phase
acquires an added fuzziness and the cell volume no longer has a
uniform value equal to the Planck constant.  Fuzzy phase space has
a direct implication for the quantization of spacetime as we will
demonstrate in section \ref{quant}.

In view of the special theory of relativity, particles are
actually located at spacetime points.  The introduction of
smearing in the spatial direction demands that we introduce
fuzziness in the time direction, otherwise, the instantaneous
annihilation of a particle of finite extent would violate
causality.  As was the case with the fuzzy position the smearing
is achieved by convolving the time coordinate with a Gaussian in
the zeroth component of the momentum operator (the Hamiltonian)
giving rise to
\bea
T_{f} &=& (Te^{-H^{2}/m^{2}})\rightarrow (te^{-H^{2}/m^{2}})\\
H &\rightarrow& i\frac{d}{dt}.
\eea
We observe that in our representation  we choose to view time as
an operator on the same footing as the position operator.  This is
in keeping with the modern unified view of spacetime and is
further evidenced when we discuss the nontrivial commutation
relations between the 4-positions.  The smeared time operator
$T_{f}$ reverts to its smooth time coordinate representation in
the limit as the characteristic times of the quantum system become
much longer than the flight time of the particle.  The time of
flight of a particle is defined as the time it takes to traverse a
distance of the Compton wavelength at the maximally allowable
speed c.  Due to the fuzziness we have introduced in the time
direction the energy-time uncertainty principle gets modified in a
manner analogous to the phase space uncertainty product giving
rise to
\be
\Delta H\Delta T \, \geq \, \frac{\hbar}{2} \,+\, \frac{2\langle
H\rangle\langle T\rangle}{m^{2}}(\Delta H)^{2}.
\ee
This relation implies a minimal uncertainty in time given by
\be
\Delta T_{0} = \frac{2}{m}\sqrt{\langle H\rangle \langle T \rangle\hbar}
\ee
which is expected since the time operator has been smeared out.
The product $\langle H\rangle\langle T\rangle$ is in general
non-negative.  It can be made negative by moving the center of the
time coordinate but this would imply that the Hamiltonian of the
underlying system obeys time translational invariance.  For all
such systems the Hamiltonian is time independent and incorporating
the time smearing into our quantum description is irrelevant to
the dynamics.  Hence, the Compton wavelength can be set to zero in
such cases which is the correspondence limit with ordinary quantum
mechanics.  Thus, by introducing these self-adjoint operator
representations for position and time we are able to quantify and
characterize the finite extent of a particle.  We now proceed to
formulate the Hilbert space representation theory of these
operators.

\section{Hilbert Space Representation}
The fuzzy position operator $X_{f}$ and the momentum operator P
satisfy the uncertainty relation Eq.~(\ref {f-eqn}).  This
relation does not imply a minimal uncertainty in the fuzzy
position or the momentum.  As a consequence, the eigenstates of
the self-adjoint fuzzy position and momentum operators can be
approximated to arbitrary precision by sequences
$|\psi_{n}\rangle$ of physical states of increasing localization
in position or momentum space:
\be
\lim_{n \rightarrow \infty}\Delta X_{f_{|\psi_{n}\rangle}} = 0 \quad
\mbox{or} \quad
\lim_{n \rightarrow \infty}\Delta P_{|\psi_{n}\rangle} = 0.
\ee
Hence, the fuzzy position and momentum operators admit a
continuous position or momentum space representation in the
Hilbert space.  Since the momentum operator is identical to the
one used in ordinary quantum mechanics it has the usual orthogonal
plane wave eigenstates.  The eigenvalue problem of the fuzzy
position operator
\be
X_{f}\psi = \lambda\psi
\ee
can be written in the momentum basis (which we choose for
convenience) as
\be
e^{-p^{2}/2m^{2}}\frac{d}{dp}(e^{-p^{2}/2m^{2}}\psi) = -i\lambda\psi.
\ee
Defining the function $\phi = e^{-p^{2}/m^{2}}\psi$ and
introducing the measure transformation $dr = e^{p^{2}/m^{2}}dp$ we
obtain the eigensolutions as
\be
\psi(p) = \frac{1}{\sqrt{2\pi}}\,e^{p^{2}/2m^{2}\,+\,i\lambda r},
\ee
where freedom in scale has been used to normalize the solution.
The eigenfunctions are orthogonal with respect to the transformed
measure $L^{2}(e^{-p^{2}/m^{2}}dr)$ because
\be
\langle \psi_{\lambda}(p)|\psi_{\lambda'}(p)\rangle = \frac{1}{2\pi}\int_{-\infty}^{\infty}e^{i(\lambda - \lambda')r}dr = \delta(\lambda - \lambda').
\ee
The inner product
$\langle\psi_{\lambda}(p)|\psi_{\lambda'}(p)\rangle$ is divergent
in the space $L^{2}(dp)$ but is equal to the Dirac delta function
in the space $L^{2}(e^{-p^{2}/m^{2}}dr)$.  As $p$ ranges from
$-\infty$ to $\infty$ the volume element $dp$, under the measure
transformation, is squeezed into a Gaussian width times the line
element $dr$, and consequently the orthogonality of the fuzzy
position eigenstates is preserved.  We note that had we tried to
construct the formal position eigenstates (eigenstates of $X$) we
would have had to sacrifice orthogonality due to the appearance of
the minimal uncertainty in position.  The eigenfunctions of the
fuzzy position operator in the position representation will be
Fourier transforms of the eigensolutions in the momentum
representation since the Fourier transform of an $L^{2}$ function
will be an $L^{2}$ function in the same measure.

\section{Translational and Rotational Invariance}
We will now examine the behavior of the quantum mechanics of
extended objects under translations and rotations and solve the
eigenvalue problem of fuzzy angular momentum.
\subsection{Translational Invariance}
Under a translation of the coordinate $x \rightarrow x + \epsilon$
we have the fuzzy translation
\bea
\langle X_{f}\rangle &\rightarrow & \langle X_{f}\rangle +
\epsilon\langle e^{-P^{2}/m^{2}}\rangle , \nonumber \\ \langle
P\rangle & \rightarrow & \langle P\rangle.
\eea
In the passive transformation picture
\bea
\label{t-eqn}
 T^{\dagger}(\epsilon)X_{f}T(\epsilon) & = &
X_{f} + \epsilon \,e^{-P^{2}/m^{2}}, \nonumber \\
T^{\dagger}(\epsilon)PT(\epsilon) &=& P,
\eea
where $T(\epsilon)$ is the translation operator which translates
the state $|\psi\rangle$.  Expanding $T(\epsilon)$ to first order
and feeding into Eq.~(\ref{t-eqn})we obtain
\be
[X_{f},G] = ie^{-P^{2}/m^{2}},
\ee
where $G$ is the generator of infinitesimal translations.  Thus,
the momentum is still the generator of fuzzy spatial translations
and analogously, we find that the Hamiltonian is the generator of
fuzzy time translations.  Since these are the same generators as
found in ordinary quantum mechanics, we can conclude by similar
reasoning and by Ehrenfest's theorem that fuzzy space (time)
translational invariance will ensure the time independence of the
momentum (Hamiltonian).

\subsection{Rotational Invariance}
Let us denote the operator that rotates two-dimensional vectors by
$R(\phi_{0}\hat{k})$ for a rotation by $\phi_{0}$ about the
z-axis.  Let $U[R]$ be the operator associated with this rotation.
For an infinitesimal rotation $\epsilon_{z}\hat{k}$ we set
\be
U[R] = I - i\epsilon_{z}L_{f_{z}},
\ee
where $L_{f_{z}}$ is the generator of fuzzy rotations.  We can
determine $L_{f_{z}} = X_{f}P_{y} - Y_{f}P_{x}$ by feeding this
$U[R]$ into the passive transformation equations for an
infinitesimal rotation:
\be
U^{\dagger}[R]X_{f}U[R] = X_{f} - Y_{f}\epsilon_{z},
\ee
and so on.  $L_{f_{z}}$ is conserved in a problem with rotational
invariance: if
\be
U^{\dagger}[R]H(X_{f},P_{x};Y_{f},P_{y})U[R] = H(X_{f},P_{x};Y_{f},P_{y})
\ee
it follows (by choosing an infinitesimal rotation) that
\be
[L_{f_{z}},H] = 0 \quad \mbox{or}\quad \langle {\dot L_{f_{z}}}\rangle = 0
\ee
by Ehrenfest's theorem.

\subsection{The eigenvalue problem of $L_{f_{z}}$}
In the momentum basis the two dimensional fuzzy angular momentum
operator can be written as
\be
L_{f_{z}} \rightarrow
e^{-p^{2}/2m^{2}}(i\frac{\partial}{\partial_{p_{x}}}e^{-p^{2}/2m^{2}}p_{y}
- i\frac{\partial}{\partial_{p_{y}}}e^{-p^{2}/2m^{2}}p_{x}),
\ee
where $p^{2} = p_{x}^{2} + p_{y}^{2}$.  This is the correct
generalization of the smeared position operator to higher
dimensions (in this case two) as can be seen by letting $X_{f}$
act on a wavefunction in two dimensions.  We can further simplify
the derivatives in $L_{f_{z}}$ and switch to polar coordinates to
obtain
\be
L_{f_{z}} \rightarrow
-ie^{-p^{2}/2m^{2}}\frac{\partial}{\partial_{p_{\phi}}}e^{-p^{2}/2m^{2}}.
\ee
The eigenvalue problem of $L_{f_{z}}$,
\be
L_{f_{z}}\psi(p_{\rho},p_{\phi}) = l_{f_{z}}\psi(p_{\rho},p_{\phi}),
\ee
can be written in the  momentum basis as
\be
-ie^{-p^{2}/2m^{2}}\frac{\partial}{\partial_{p_{\phi}}
}(\psi e^{-p^{2}/2m^{2}}) = l_{f_{z}}\psi.
\ee
Defining $\phi = \psi e^{-p^{2}/m^{2}}$ and using the transformed measure,
\be
dp_{\phi} = \frac{1}{2\pi}[\frac{\sqrt{\pi}m}{2i}erf(2\pi i)]
e^{-p_{\phi}^{2}/m^{2}}\, dr
\ee we arrive at
\be
\psi(p_{\rho},p_{\phi}) \,\sim\,
e^{il_{f_{z}}e^{p_{\rho}^{2}/m^{2}}r\, +\, p^{2}/2m^{2}},
\ee
where the numerical factor in the measure transformation has been
chosen so that as $p_{\phi}$ ranges from 0 to $2\pi$,
$r$ also ranges from 0 to $2\pi$. The eigenfunctions are orthogonal
with respect to the transformed measure
$L^{2}(e^{-p_{\phi}^{2}/m^{2}}p_{\rho}dp_{\rho}dr)$ where the
numerical factor has been suppressed.  We  observe that
$l_{f_{z}}$ seems to be arbitrary and even complex since the range
of $r$ is restricted.  The fact that complex eigenvalues enter the
solution signals that we are overlooking the Hermiticity
constraint.  Imposing this condition we have
\be
\langle \psi_{1}|L_{f_{z}}|\psi_{2}\rangle =  \langle
\psi_{2}|L_{f_{z}}|\psi_{1}\rangle^{*},
\ee
which becomes in the momentum basis
\be
\int_{0}^{\infty}\int_{0}^{2\pi}
\phi_{1}^{*}(-i\frac{\partial}{\partial_{p_{\phi}}})\phi_{2}\,
p_{\rho}dp_{\rho}dp_{\phi} =
\left[\int_{0}^{\infty}\int_{0}^{2\pi}
\phi_{2}^{*}(-i\frac{\partial}{\partial_{p_{\phi}}})\phi_{1}\,
p_{\rho}dp_{\rho}dp_{\phi}\right]^{*},
\ee
where $\phi = \psi e^{-p^{2}/2m^{2}}$.  If this requirement is to
be satisfied by all $\phi_{1}$ and $\phi_{2}$, one can show (by
integrating by parts) that it is enough if each
$\phi(p_{\rho},p_{\phi})$ obeys
\be
\phi(p_{\rho},0) = \phi(p_{\rho},2\pi).
\ee
If we impose this constraint on the $L_{f_{z}}$ eigenfunctions we
find that the eigenvalues $l_{f_{z}}$ have to obey the following
relation
\be
l_{f_{z}} = e^{-p_{\rho}^{2}/m^{2}}k,
\ee
where $k$ is an integer.  The fuzzy angular momentum is equal to
an integral multiple of $\hbar$ times a smearing factor.  This is
an example of smeared or fuzzy quantization and as the Compton
wavelength vanishes we regain the usual relation for ordinary
quantized angular momentum.

\section{Quantization of Spacetime}
\label{quant} The raised phase space uncertainty product which we
have discussed before implies that phase space acquires an added
fuzziness due to the smearing of the position operator.  By
considering the algebra of smooth functions over fuzzy phase space
generated by fuzzy positions and momenta, and by using the
Gel'fand and Naimark reconstruction theorem one can recover all
information about the underlying space.  However, since we already
know the mathematical form of the fuzzy position operator, we use
a more simple approach and directly construct the nontrivial
commutators between the fuzzy positions.  In the momentum basis
the commutator between fuzzy positions in 4-dimensional spacetime
is
\be
\left[X_{f_{\mu}},X_{f_{\nu}}\right] =
-e^{-p^{2}/2m^{2}}(\partial_{p_{\mu}}e^{-p^{2}/m^{2}}\partial_{p_{\nu}}
-
\partial_{p_{\nu}}e^{-p^{2}/m^{2}}\partial_{p_{\mu}})e^{-p^{2}/2m^{2}}.
\ee
The derivative terms can be further simplified and introducing
$X_{\mu} \rightarrow i\partial_{p_{\mu}}$ and $P \rightarrow p$ we
obtain
\be
\label{xf-eqn}
\left[X_{f_{\mu}},X_{f_{\nu}}\right] =
\frac{i}{m^{2}}e^{-P^{2}/2m^{2}}(P_{\nu}X_{\mu} -
P_{\mu}X_{\nu})e^{-P^{2}/2m^{2}}.
\ee
The nontrivial commutation relation between the fuzzy positions implies
that fuzzy spacetime is quantized.  When the confines are much larger
than the Compton wavelength, that is, when we are viewing a larger patch
of spacetime, ${p^{2}}/{m^{2}} \ll 1$, and
the Gaussian (smearing) factors in Eq.~(\ref{xf-eqn}) become negligible.
In this limit  $X_{f_{\mu}} \rightarrow X_{\mu}$, and we obtain
\be
\label{q-eqn}
[X_{f_{\mu}},X_{f_{\nu}}] \rightarrow
[X_{\mu},X_{\nu}] = \frac{i}{m^{2}}(P_{\nu}X_{\mu} -
P_{\mu}X_{\nu}).
\ee
Thus, as long as the Compton wavelength is nonzero, the ordinary
4-positions also exhibit a nontrivial commutation relation given
by Eq.~(\ref{q-eqn}).This result is identical to the one obtained
by Snyder in 1947\cite{snyder}.  In his paper Snyder demonstrates
that the assumption of Lorentz covariance does not exclude a
quantized spacetime which he develops by defining the 4-positions
in terms of the homogenous (projective) coordinates of a De Sitter
space.  In the limit as the natural unit of length (the Compton
wavelength) vanishes our quantized spacetime changes to the
ordinary continuous spacetime and the commutators revert to their
standard values.  Therefore, our formulation of the quantum
mechanics of extended objects implies that spacetime is quantized
and that it has a Lorentz covariant structure.

\section{Fuzzy (extended object) Harmonic Oscillator}
Before we study the quantum mechanical fuzzy harmonic oscillator
let us understand the classical analog of such an oscillator.
Classically, we can model an extended object as a point mass
connected to a nonlinear spring of stiffness constant, say
$k_{1}$.  When this spring-mass system is connected to another
linear spring of stiffness constant, say $k_{2}$ we essentially
have a classical, one dimensional, extended object oscillator.
When the wavelength of oscillation is small compared to the size
of the extended object (in this case the length of the nonlinear
spring of stiffness constant $k_{1}$) the oscillator will exhibit
harmonic behavior since the small oscillations do not disturb the
configuration of the extended object.  As the wavelength of
oscillation becomes comparable to the size of the extended object,
anharmonic vibrations set in.  Again, as the wavelength of
oscillation becomes much larger than the size of the extended
object, the point particle approximation becomes tenable and
harmonic vibrations are recovered.  We would expect the quantum
version of the extended object oscillator to exhibit similar
behavior albeit with quantized energy levels.  In the first
regime, when the wavelength of oscillation is small compared to
the size of the extended object, since small oscillations do not
disturb the configuration of the extended object to any
appreciable extent we will obtain the usual quantized energy
levels of the simple harmonic oscillator.  It is in the second and
third regimes where we would need to apply the quantum mechanics
of extended objects.  The Hamiltonian for a one dimensional fuzzy
harmonic oscillator can be written as
\be
H = \frac{P^{2}}{2m} + \frac{1}{2}m\omega^{2}X_{f}^{2}.
\ee
Introducing the operator representation for the fuzzy position and
momentum in the momentum basis and simplifying terms, we obtain
\be
\label{de-eqn} \frac{1}{2}m\omega^{2}
\left[\frac{d^{2}\phi}{dp^{2}} - (\frac{p^{2}}{m^{4}} -
\frac{1}{m^{2}})\phi\right]
    = (\frac{p^{2}}{2m} - E)e^{2p^{2}/m^{2}}\phi,
\ee
where $\phi = e^{-p^{2}/m^{2}} \psi$, $H\psi = E\psi$, and $\phi$
lies in $L^2(dp)$. When the wavelength of oscillation (the
confines) is large compared to the size of the extended object,
${p^{2}}/{m^{2}} \ll 1$, in which case we can approximate
$e^{2p^{2}/m^{2}} \approx 1 + {2p^{2}}/{m^{2}}$. In this
approximation Eq.~(\ref{de-eqn}) can be rewritten as:
\be
\frac{d^{2}\phi}{dp^{2}} + 2m({\tilde E} -\frac{1}{2}m\Omega^{2})\phi = 0,
\ee
where
\bea
2m{\tilde E} &=& \frac{2E}{m\omega^{2}} + \frac{1}{m^{2}}, \\
m^{2}\Omega^{2} &=&\frac{-4E}{m^{3}\omega^{2}} +
\frac{1}{m^{4}} + \frac{1}{m^{2}\omega^{2}}.
\eea
This is simply the differential equation for a simple harmonic
oscillator in terms of the dummy energy ${\tilde E}$ and frequency
$\Omega$.  For well behaved solutions we require the quantization
condition
\be
{\tilde E}_{n} = (n + \frac{1}{2})\Omega,\; n = 0,1,2,\ldots.
\ee
Re-expressing this relation in terms of the physical energy $E$
and frequency $\omega$ and retaining terms up to $o(\hbar^{2})$,
we obtain
\be
\label{es-eqn}
E_{n} = (n + \frac{1}{2})\omega - \frac{\omega^{2}}{2m},\; n = 0,1,2,\ldots.
\ee
As we would expect, the fuzzy particle exhibits harmonic behavior
when the wavelength of oscillation is large compared to the size
of the particle.  In this approximation, the eigenvalue spectrum
of the fuzzy harmonic oscillator is equivalent to the spectrum of
a displaced simple harmonic oscillator.  The shift in the energy
spectrum can be understood by observing that in the classical
spring-mass model, the extended object (the nonlinear spring)
would undergo compression due to the oscillations of the linear
spring thereby displacing the equilibrium position.  The quantum
counterpart exhibits the same behavior and when $\omega \ll m$ in
Eq.~(\ref{es-eqn}), that is, when the point particle approximation
becomes tenable we obtain the eigenspectrum of the simple harmonic
oscillator.  In the classical analog this would mean that, at
sufficiently large oscillation wavelengths the compression of the
nonlinear spring becomes insignificant.  Retaining terms up to
$o(\hbar^{2})$, the eigenfunctions of the harmonic oscillator in
this approximation are given by:
\be
\psi(p)\,\sim\, e^{p^{2}/m^{2}(1 -
\frac{m}{2\omega})}H_{n}[\sqrt{{(m\omega)^{-1}}}p],
\ee
where $H_{n}$ are the Hermite polynomials.  Since $\psi$ lies in
$L^2(e^{-2p^2/m^2}dp)$, the eigenfunctions will be
normalizable.
  By
inserting these approximate solutions into the exact differential
equation Eq.~(\ref{de-eqn}) we find that they do not differ by
derivative terms and hence they are close in some sense to the
exact solutions.

If we include higher values of momenta in our approximation and write
$
e^{2p^{2}/m^{2}} \approx 1 + {2p^{2}}/{m^{2}} + {2p^{4}}/{m^{4}},
$
we obtain the differential equation
\be
\frac{d^{2}\phi}{dp^{2}} + 2m(\frac{\alpha}{2m} -
\frac{\beta}{2m}p^{2} - \frac{\gamma}{2m}p^{4})\phi = 0,
\ee
where
\bea
\alpha &=& \frac{2E}{m\omega^{2}} + \frac{1}{m^{2}}, \\
\beta &=&
\frac{-4E}{m^{3}\omega^{2}} + \frac{1}{m^{4}} +
\frac{1}{m^{2}\omega^{2}}, \\ \gamma &=& \frac{2}{m^{4}\omega^{2}}
- \frac{4E}{m^{5}\omega^{2}}.
\eea
This is the differential equation for an anharmonic oscillator.
As we would expect when higher momentum values become important or
equivalently as the wavelength of oscillation becomes comparable
to the size of the fuzzy particle, anharmonic vibrations set in.
We can compute the eigenspectrum of the anharmonic oscillator
using perturbation theory.  We note that the perturbation
expansion breaks down for some large enough $n$.  Retaining terms
up to $o(\hbar^{2})$ the eigenspectrum is found to be
\be
E_{n} = (n + \frac{1}{2})\omega - \frac{\omega^{2}}{2m} +
\frac{3\omega^{2}}{4m}(1 + 2n + 2n^{2}), \; n = 0,1,2,\ldots.
\ee
Figure 1 shows a plot of the first two anharmonic oscillator
eigenfunctions.  For comparison the first two harmonic oscillator
eigenfunctions are also shown.  The anharmonic oscillator
eigenfunctions have a steeper slope because the particle is placed
in a stronger potential as compared to the harmonic oscillator
potential.  If we include even higher values of momenta in our
approximation we find that the anharmonicity increases and in the
limit of large quantum numbers our quantum descriptions pass
smoothly to their classical counterparts.  Therefore, the quantum
mechanics of extended objects provides a description of the fuzzy
harmonic oscillator which augments our classical intuition.  Such
a description could be useful when we study harmonic excitations
of quasiparticles which cannot be localized to arbitrary
precision.  The quantum mechanics of extended objects can also be
used to describe compound particles such as baryons or mesons in
situations where their nonzero size matters but the details of the
internal structure do not contribute.  One such situation is the
description of the nucleon-nucleon interaction at very short
distances which we proceed to examine.

\section{The Yukawa Potential}
At present the physics of the nucleon-nucleon interaction can be
divided into three major regions\cite{weise}
\begin{enumerate}
\item The {\it long-distance} region $r \geq 2$ fm $\approx
1.5m_{\pi}^{-1}$ where one-pion exchange dominates and the
quantitative behavior of the potential is very well established;
\item   The {\it intermediate} region $0.8$ fm $\leq r \leq 2$ fm
where the dynamical contributions from two-pion exchange
(effective boson exchange) compete with or exceed the one-pion
exchange potential;
\item The {\it inner} region $r \leq 0.8$ fm
has a complicated dynamics not readily accessible to a
quantitative theoretical description.  This region is expected to
be influenced by heavy mesons and or by quark/gluon degrees of
freedom.  It is usually approached in a phenomenological way.
\end{enumerate}
Moreover, the inner region contains a repulsive hard core of
radius $0.6$ fm which was first proposed by Jastrow in 1951 in
order to fit nucleon-nucleon scattering data\cite{jastrow}.  The
presence of a repulsive nucleon core is necessary to explain the
saturation of nuclear forces.  This short range and repulsive
nucleon force is believed to be mediated by an $\omega$ meson of
mass $782$ MeV and the intermediate range attractive nucleon force
is mediated by a $\sigma$ meson (effective boson) of mass $550$
MeV\cite{walecka}.  Once the masses are fixed, the coupling
constants which measure the strength of the coupling between a
meson and a baryon are chosen to reproduce nucleon-nucleon
scattering phase shifts and deuteron properties.  These
phenomenological coupling constants\cite{walecka} are found to be
${g_{\omega}^{2}}/{4\pi} = 10.83$ and ${g_{\sigma}^{2}}/{4\pi} =
7.303$.  It is our objective to theoretically determine the radius
of the repulsive nucleon core and to reproduce the
phenomenological $\omega$ meson coupling constant using the
quantum mechanics of extended objects which becomes relevant to
the dynamics in the inner region due to the finite extent of the
nucleon.

In order to reproduce consistent results we will focus attention
on the bound state nucleon-nucleon interaction, namely, the
deuteron.  The deuterium nucleus ($A = 2, Z = N = 1$) is a bound
state of the neutron-proton system, into which it may be
disintegrated by irradiation with $\gamma$ rays of energy above
the binding energy\cite{sachs} of $2.226$ MeV.  The ground state
of the deuteron is a triplet $S$ state and it has no excited
states.  The force between the proton and the neutron can be
described in good approximation by a potential energy function of
the form
\be
V(r) = -V_{0}\frac{e^{-r/r_{0}}}{r/r_{0}}.
\ee
This is the well known Yukawa potential and is central to the
mesonic theory of nuclear forces.  The range of the force $r_{0}$
is equal to ${1}/{\mu}$, where $\mu$ is the mass of the associated
meson and the strength $V_{0}$, or depth of the potential well is
connected with the strength of the coupling between the meson and
the nucleon field.  In the center-of-mass coordinates the
Hamiltonian for the $S$ state of the deuteron is
\be
H = \frac{p^{2}}{2m} +V(r),
\ee
where $m$ is the reduced mass of the deuteron and $r$ determines
the neutron-proton separation.  For ease of comparison with the
quantum mechanics of extended objects in which the momentum basis
is more convenient, we can transcribe the Hamiltonian to the
momentum basis by virtue of the exchange transformation
\be
r \rightarrow pr_{0}^{2}, \quad {\rm and } \quad p \rightarrow -r/r_{0}^{2}.
\ee
The exchange transformation is a canonical transformation and does
not affect the dynamics\cite{goldstein}.  The Hamiltonian in the
momentum basis is
\be
H = \frac{r^{2}}{2mr_{0}^{4}} + V(p),
\ee
where ${\bf r} \rightarrow i\nabla_{\bf p}$ is the position operator and
$V(p) = -V_{0}{e^{-pr_{0}}}/{pr_{0}}$.  The binding energy
$E_{0} = -2.226$ MeV can be estimated by means of the variational principle
using the simple trial wavefunction
\be
\psi(p) = e^{-\alpha pr_{0}},
\ee
in which we treat $\alpha$ as a variable parameter.  Our choice of
the trial wavefunction is motivated by the fact that we expect the
ground state wavefunction to have no angular momentum, no nodes,
and for $p\psi(p)$ to vanish as $p \rightarrow \infty$ as required
for bound states.  The variational method determines the energy as
\be
E = \frac{\langle\psi|H|\psi\rangle}{\langle\psi|\psi\rangle}.
\ee
The energy $E$ serves as an upper bound on the ground state energy $E_{0}$.  If we substitute $E_{0} = -2.226$ MeV for $E$ we can perform an approximate calculation of the relation between $V_{0}$ and $r_{0}$ (range-depth relation) that must hold if the potential function $V(p)$ is to give the value $E_{0} = -2.226$ MeV for the binding energy.  Figure $2$ shows a plot of the range-depth relation for the Yukawa potential (deuteron) as determine by this method.  By comparing the values of $V_{0}$ for various values of $r_{0}$ with the results of an exact calculation using numerical integration we are able to estimate the accuracy of our approximate result.  The approximate result is within a few percent of the exact result and the error decreases with increasing $r_{0}$\cite{sachs}.  Therefore, our choice of the trial wavefunction is justified.

Let us now analyze the same potential problem using the quantum
mechanics of extended objects.  In the momentum basis the fuzzy
Hamiltonian for the $S$ state of the deuteron is
\be
H = \frac{r_{f}^{2}}{2mr_{0}^{4}} + V(p),
\ee
where
\be
{\bf r}_f \rightarrow i\, e^{-p^{2}/2m^{2}}\nabla_{\bf p}e^{-p^{2}/2m^{2}}
\ee
is the fuzzy position operator which now determines the
neutron-proton separation.  Figure $3$ shows a plot of the $S$
state eigenfunctions as a function of momentum for $r_{0} = 1.43$
fm, which correspond to a $\pi$ meson of mass $139.6$ MeV, and for
$r_{0} = 0.3596$ fm, which corresponds to a $\sigma$ meson of mass
$550$ MeV.  The eigenfunctions obtained from ordinary quantum
mechanics are also shown for comparison.  The eigenfunctions
obtained from the quantum mechanics of extended objects are pushed
out in comparison to the usual eigenfunctions implying that there
is a repulsive component to the potential which has the effect of
pushing out the eigenfunctions as at the edge of an infinite well
(compare with figure 1).  By examining the plots of $\phi(p) =
e^{-p^{2}/m^{2}}\psi(p)$ (figure 4 shows one such plot for $r_{0}
= 1.43$ fm) where $\psi(p)$ are the eigenfunctions obtained from
the quantum mechanics of extended objects,  we observe that
$\phi(p)$ lies in $L^{2}(d^{3}p)$.  Therefore, the eigenfunctions
obtained from the extended object analysis are  normalizable with
respect to $L^{2}(e^{-2p^{2}/m^{2}}d^{3}p)$.  This motivates us to
choose as our trial wavefunction
\be
\label{trial}
\psi(p) = e^{p^{2}/m^{2} - \alpha pr_{0}}.
\ee
The normalizability criterion in this measure ensures that
\be
e^{-p^{2}/m^{2}}p\psi(p) \rightarrow 0 \mbox{ as } p \rightarrow \infty
\ee
as required for bound states (and as is the case with our trial
wavefunction).  Furthermore, when the confines are large
($p^{2}/m^{2} \ll 1$), $\psi(p)$ in Eq.~(\ref{trial}) passes
smoothly into the trial wavefunction we had used when we applied
ordinary quantum mechanics and which had yielded an accurate
range-depth relation.  Hence, our choice of the trial wavefunction
is justified and with the given volume element we can determine
the approximate range-depth relation that must hold if the
potential function $V(p)$ is to give the value $E_{0} = -2.226$
MeV for the binding energy.  Numerical calculations performed in
Mathematica reveal the range-depth relation shown in figure 5. The
strength of the potential or depth of the well $V'_{0}$ in figure
5 is lower than the strength of the potential $V_{0}$ obtained
from ordinary quantum mechanics (figure 2) particularly for
smaller values of $r_{0}$.  The existence of a repulsive component
to the potential which we have already observed from a plot of the
eigenfunctions shown in figure 3 is verified. Moreover, the depth
of the well $V'_{0}$ in figure 5 is negative for $r_{0} \leq
0.563$ fm.  This implies the existence of a repulsive nucleon core
with a radius $r_{c} = 0.563$ fm, which is consistent with the
phenomenologically obtained value of $0.6$ fm.

Let us model the effective nucleon-nucleon interaction by a
potential of the form
\be
\label{nn-eqn}
V(r) = -V_{0}\frac{e^{-r/r_{0}}}{r/r_{0}} +
V_{1}\frac{e^{-r/r_{1}}}{r/r_{1}},
\ee
where $r_{0} = 0.3596$ fm corresponding to $\sigma$ meson exchange
(attraction)and $r_{1} = 0.2529$ fm corresponding to $\omega$
meson exchange (repulsion).  This potential describes the main
qualitative features of the nucleon-nucleon interaction: a short
range repulsion between baryons coming from $\omega$ exchange and
an intermediate range attraction coming from $\sigma$
exchange\cite{walecka}.  The repulsive component of the effective
nucleon-nucleon interaction must be held accountable for the drop
in the well depth from $V_{0}$ to $V'_{0}$, which is observed at
$r_{0} = 0.3596$ fm.  Since the $\omega$ exchange occurs at a
range of $r_{1} = 0.2529$ fm we require that
\be
V(r = r_{1}) = -V'_{0}\,\frac{e^{-r_{1}/r_{0}}}{r_{1}/r_{0}}.
\ee
The quantities $V_{0} = 660.77$ MeV and $V'_{0} = -81.0$ MeV can
be computed numerically or can be read from figures 2 and 5. A
simple calculation yields the strength of the repulsive potential
as $V_{1} = 1419.07$ MeV.  Figure 6 shows a plot of the effective
nucleon-nucleon interaction.  The potential is attractive at large
distances and repulsive for small $r$.  In terms of the coupling
constants we can rewrite the effective nucleon-nucleon interaction
as
\be
V(r) = \frac{-g_{\sigma}^{2}}{4\pi}\,\frac{e^{-r/r_{0}}}{r} +
\frac{g_{\omega}^{2}}{4\pi}\,\frac{e^{-r/r_{1}}}{r}.
\ee
Comparison with Eq.~(\ref{nn-eqn}) yields ${g_{\sigma}^{2}}/{4\pi}
= 1.20$ and ${g_{\omega}^{2}}/{4\pi} = 1.815$.  Note that we are
working in units with $\hbar = c =  1$.  These theoretically
obtained values of the coupling constants will differ from the
phenomenological coupling constants because in our simple Yukawa
model of the effective nucleon-nucleon interaction we have
neglected important tensor interactions and spin-orbit terms which
contribute to the form of the potential\cite{weise}.  However, the
ratio of the theoretical coupling constants
${g_{\omega}^{2}}/{g_{\sigma}^{2}} = 1.512$ which compares the
relative strength of the repulsive coupling and the attractive
coupling must be equal to the ratio of the phenomenologically
determined coupling constants
${g_{\omega_{p}}^{2}}/{g_{\sigma_{p}}^{2}}$ in order for our
simple Yukawa model to successfully describe the effective
nucleon-nucleon interaction and to ensure the stability of the
deuteron. Using the value ${g_{\sigma_{p}}^{2}}/{4\pi} = 7.303$
and multiplying by the ratio 1.512 we obtain the value of the
phenomenological coupling constant of the $\omega$ meson as
${g_{\omega_{p}}^{2}}/{4\pi} = 11.03$.  This value of the coupling
constant differs by $1.85$ percent from the value obtained from
fitting the nucleon-nucleon scattering phase shifts and deuteron
properties which is equal to $10.83$.  Therefore, the quantum
mechanics of extended objects  leads us to values of the $\omega$
meson coupling constant and of the repulsive core radius which are
consistent with the phenomenologically obtained values.

\section{Conclusion}
In this paper we have developed the Hilbert space representation
theory of the quantum mechanics of extended objects and applied it
to the fuzzy harmonic oscillator and the Yukawa potential.  The
results of the fuzzy harmonic oscillator are consistent with our
classical intuition and in the case of the Yukawa potential we
obtain accurate theoretical predictions of the hitherto
phenomenologically obtained nucleon core radius and the $\omega$
meson coupling constant.  In an age of increasing miniaturization,
it is conceivable that as the confines of various quantum systems
become comparable to the finite extent of the confined particles,
the quantum mechanics of extended objects will play an important
role in determining the dynamics.  Furthermore, the infinite
dimensional generalization of the quantum mechanics of extended
objects, namely, the quantum field theory of extended objects
needs to be understood.  Since the ubiquitous and troublesome
vertex in quantum field theory is effectively smeared out in such
a treatment, it is possible that the problem of nonrenormalizable
quantum field theories can be rendered tractable.  The author is
pursuing investigations in this direction.

\vspace{1in}

\centerline{Acknowledgements}
I would like to thank E.C.G.~Sudarshan and L.~Sadun for insightful
discussions.  I would also like to thank R. Zgadzaj for helping
me with the numerical calculations in Mathematica.


\newpage
\begin{figure}
\caption{
The first two eigenfunctions of the anharmonic 
oscillator(solid curves). For comparison the first two 
eigenfunctions of the anharmonic oscillator are also 
shown(dashed curves). The anharmonic oscillator eigenfunctions 
show a steeper slope because the particle experiences a 
stronger potential.\label{aho}}
\end{figure}

\begin{figure}
\caption{The range-depth relation obtained by using a variational 
approximation.\label{v0-r0}}
\end{figure}

\begin{figure}
\caption{The  solid curves show the eigenfunctions obtained from 
an extended object analysis. The top figure shows the eigenfunction 
at a range of $r_{0} = 1.43fm$ and the bottom figure shows the 
eigenfunction at a range of $r_{0} = 0.3596fm$.  For comparison 
the eigenfunctions obtained from ordinary quantum mechanics are 
also shown (dashed curves). The repulsion experienced by the nucleons, 
which is important at short distances, has the effect of pushing 
out the eigenfunctions.\label{e-fns}}
\end{figure}

\begin{figure}
\caption{The plot of $\phi(p) = e^{-p^{2}/m^{2}}\psi(p)$ where 
$\psi(p)$ is the wavefunction obtained from the quantum mechanics 
of extended objects at $r_{0} = 1.43fm$.\label{phi}}
\end{figure}

\begin{figure}
\caption{The range-depth relation obtained from an extended 
object analysis using the variational method.  The strength 
of the potential is lowered, particularly for smaller values 
of $r_{0}$, indicating the existence of a repulsive component 
to the potential.\label{v0'-r0}}
\end{figure}

\begin{figure}
\caption{The effective nucleon-nucleon interaction.\label{nn}}
\end{figure}

\newpage
\begin{figure}
\centerline{
\epsfxsize=6.5in
\epsfbox{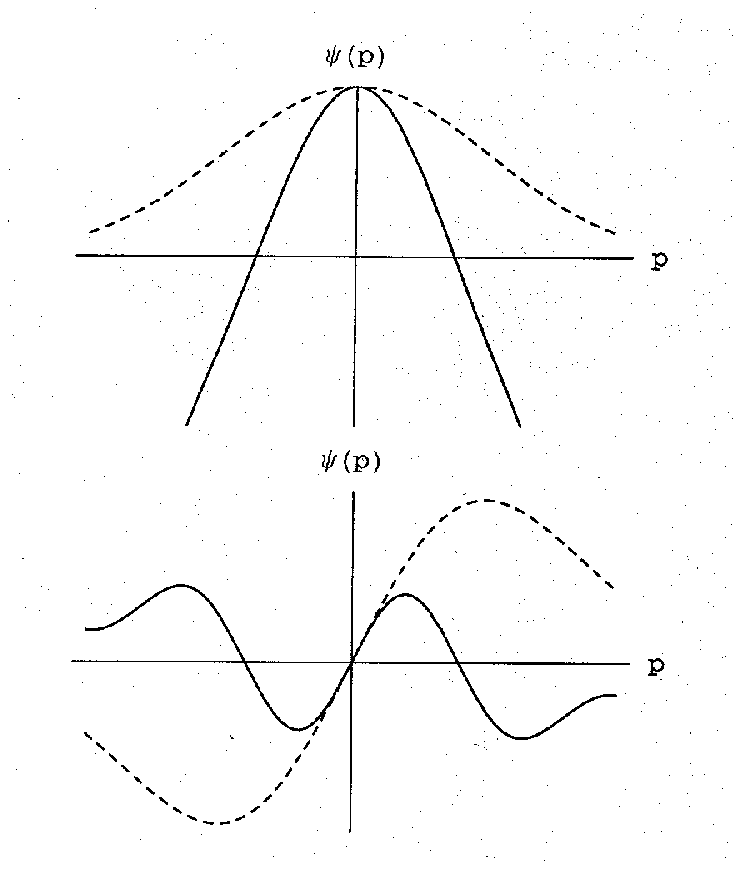}}
\end{figure}

\newpage
\begin{figure}
\centerline{
\epsfxsize=6.5in
\epsfbox{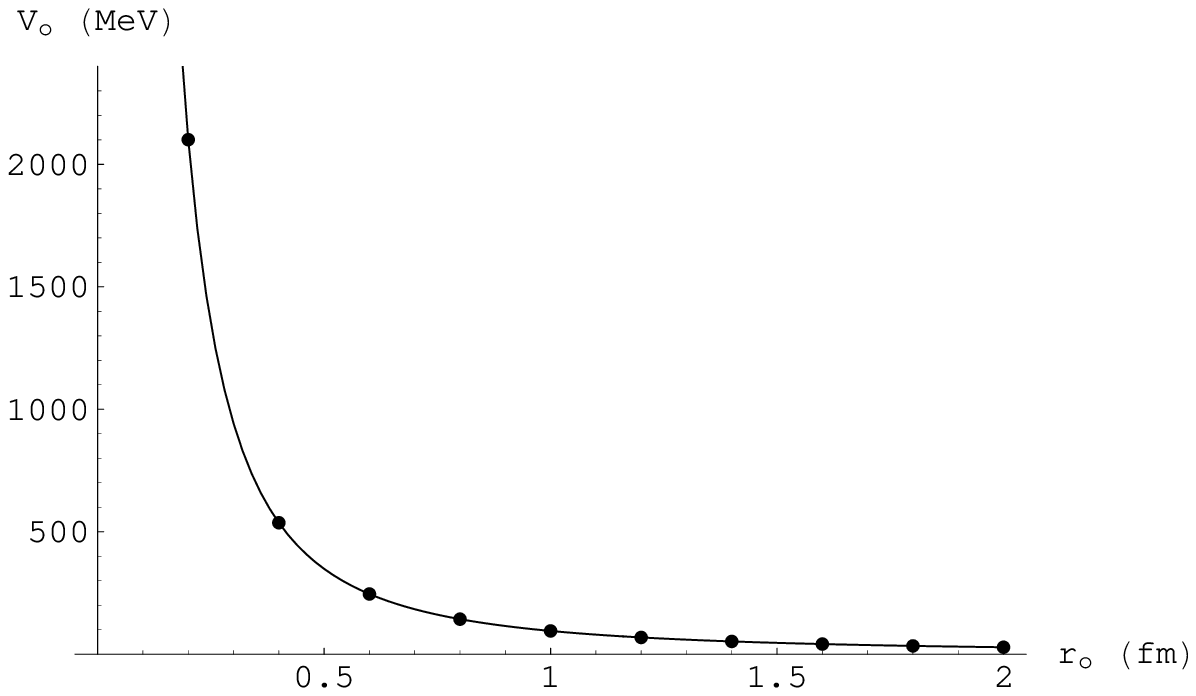}}
\end{figure}

\newpage
\begin{figure}
\centerline{
\epsfxsize=6.5in
\epsfbox{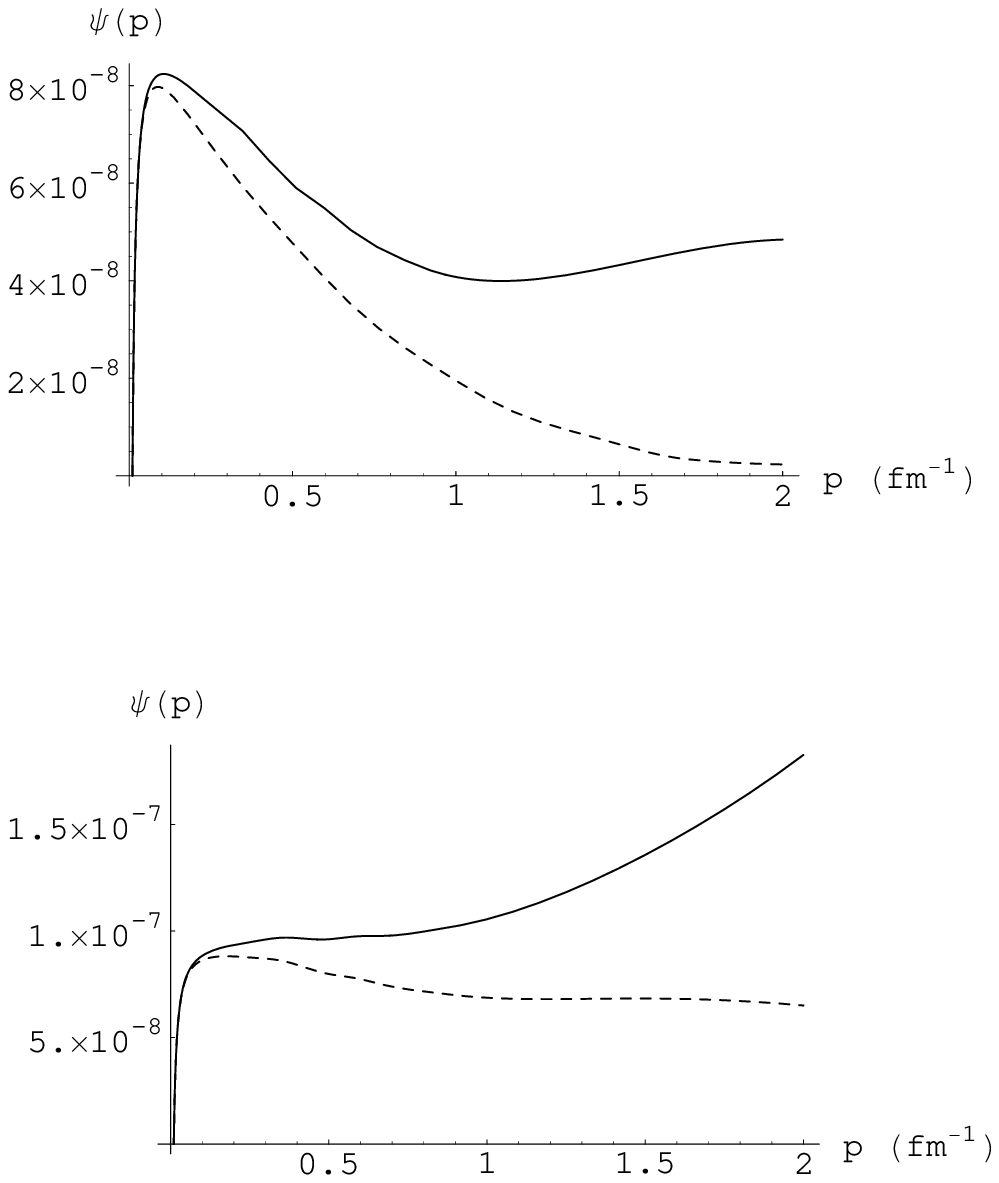}}
\end{figure}

\newpage
\begin{figure}
\centerline{
\epsfxsize=6.5in
\epsfbox{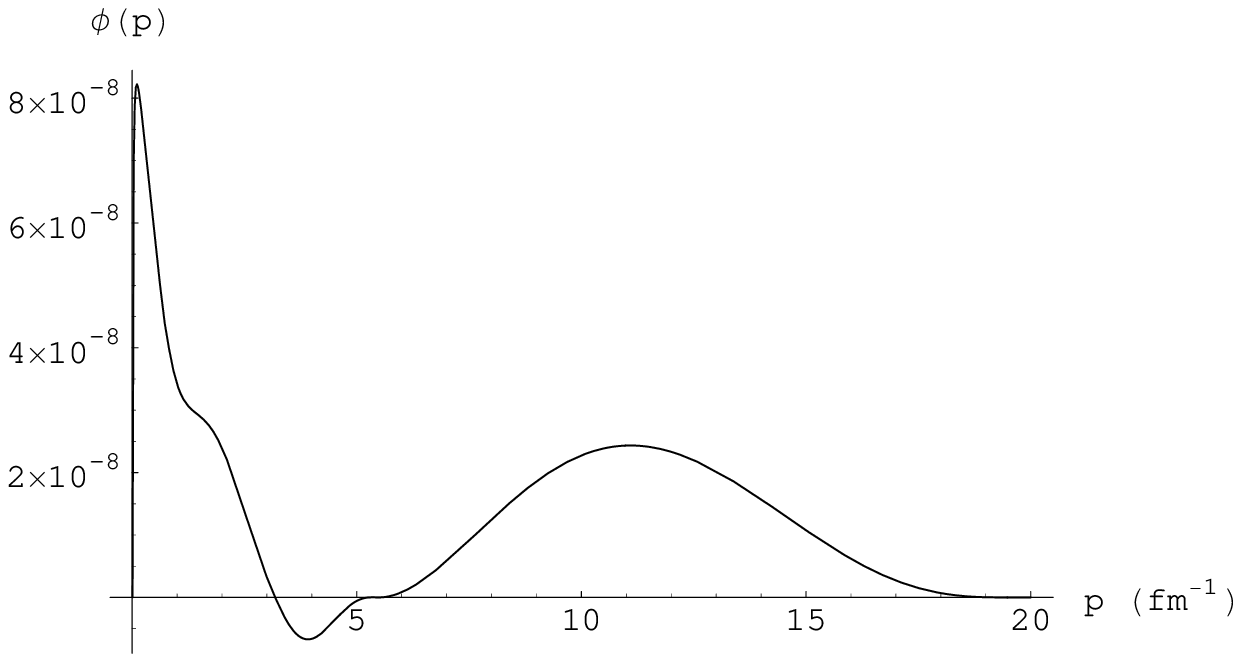}}
\end{figure}

\newpage
\begin{figure}
\centerline{
\epsfxsize=6.5in
\epsfbox{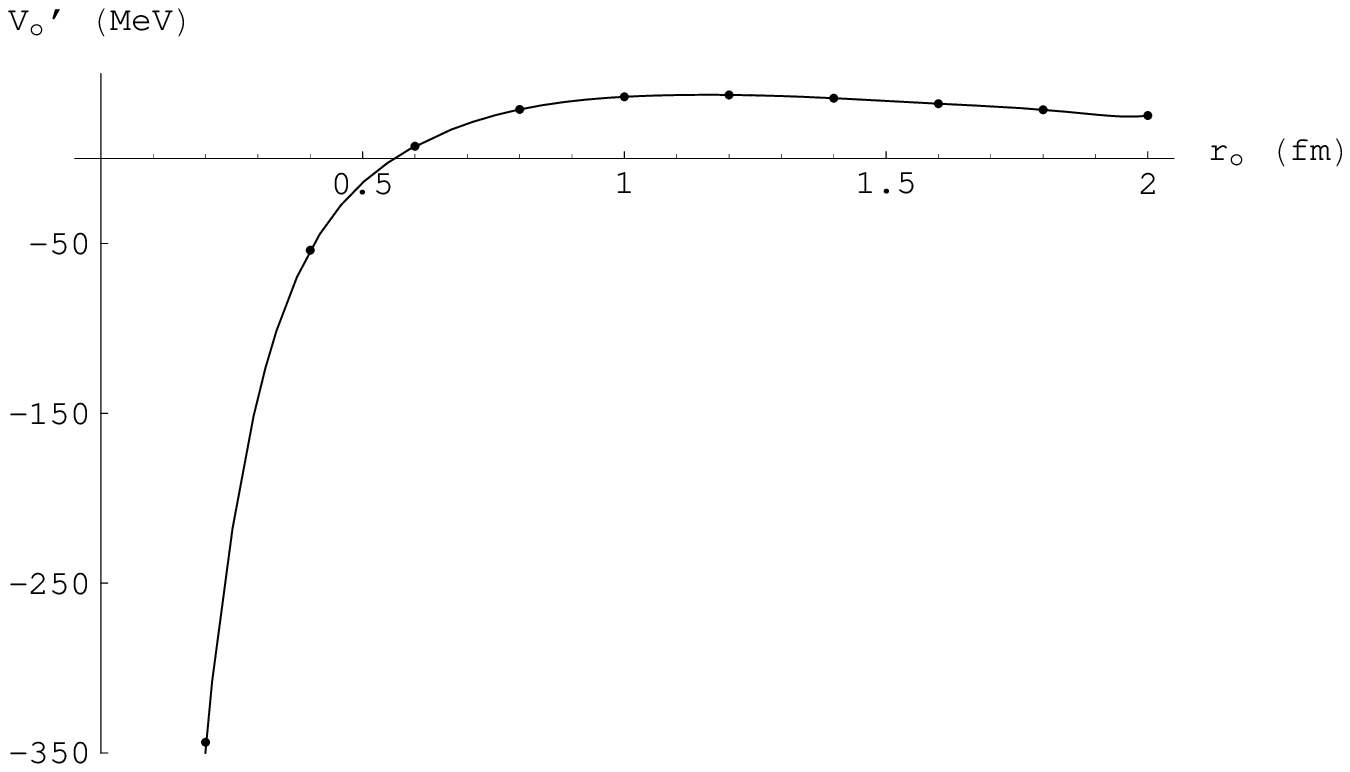}}
\end{figure}

\newpage
\begin{figure}
\centerline{
\epsfxsize=6.5in
\epsfbox{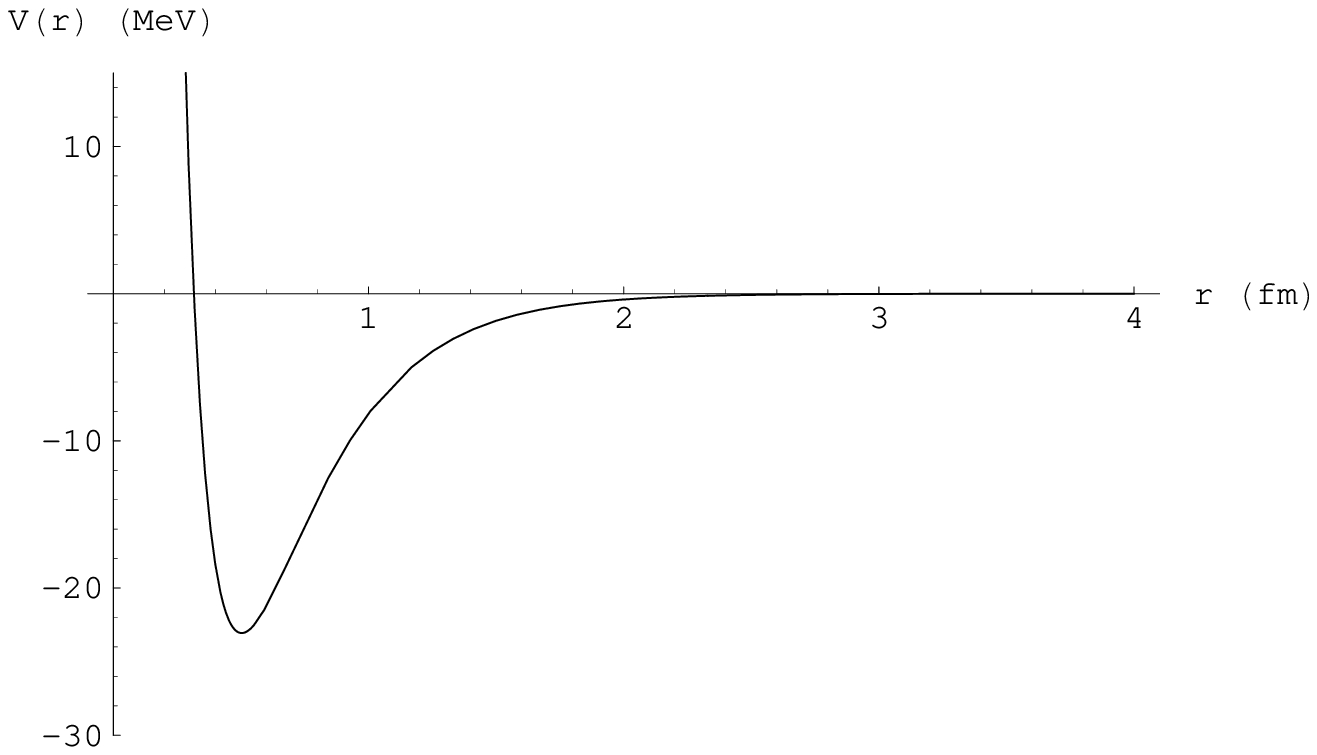}}
\end{figure}

\end{document}